\documentclass{JHEP3}

\usepackage{epsfig,multicol,bbm}

\newcommand\fverb{\setbox\fverbbox=\hbox\bgroup\verb}
\newcommand\fverbdo{\egroup\medskip\noindent%
			\fbox{\unhbox\fverbbox}\ }
\newcommand\fverbit{\egroup\item[\fbox{\unhbox\fverbbox}]}
\newbox\fverbbox

\title{Post-Newtonian cosmological dynamics of plane-parallel perturbations and back-reaction}

\author{Eleonora Villa \\ Dipartimento di Fisica, Universit\`a di Milano, via Celoria 16, 20154 Milano, Italy \\ Email: \email{eleonora.villa@unimi.it} }
\author{Sabino Matarrese \\ Dipartimento di Fisica "G. Galilei", Universit\`a degli Studi di Padova and INFN Sezione di Padova, via Marzolo 8, 35131 Padova, Italy \\ Email: \email{sabino.matarrese@pd.infn.it}} 
\author{Davide Maino \\ Dipartimento di Fisica, Universit\`a di Milano, via Celoria 16, 20154 Milano, Italy \\ Email: \email{davide.maino@mi.infn.it} }

\abstract{We study the general relativistic non-linear dynamics of self-gravitating irrotational dust in a cosmological 
setting, adopting the comoving and synchronous gauge, where all the equations can be written in terms of the 
metric tensor of spatial hyper-surfaces orthogonal to the fluid flow. Performing an expansion in 
inverse powers of the speed of light, we obtain the post-Newtonian equations, which yield the lowest-order
relativistic effects arising during the non-linear evolution.
We then specialize our analysis to globally plane-parallel configurations, 
i.e. to the case where the initial perturbation field depends on a single coordinate. 
The leading order of our expansion, corresponding to the ``Newtonian background", is the 
Zel'dovich approximation, which, for plane-parallel perturbations in the Newtonian limit, represents an exact solution.
This allows us to find the exact analytical form for the post-Newtonian metric, thereby providing
the post-Newtonian extension of the Zel'dovich solution: this accounts for
some relativistic effects, such as the non-Gaussianity of primordial perturbations. 
An application of our solution in the context of the back-reaction proposal is eventually given, 
providing a post-Newtonian estimation of kinematical back-reaction, mean spatial curvature,
average scale-factor and expansion rate.}

\dedicated{}
\renewcommand{\a}{\alpha}
\renewcommand{\b}{\beta}
\renewcommand{\d}{\delta}
\renewcommand{\gg}{\gamma}
\renewcommand{\r}{\varrho}
\renewcommand{\ss}{\sigma}

\renewcommand{\t}{\Theta} 

\newcommand{\D}{\mathcal{D}}
\newcommand{\av}[1]{\langle{#1}\rangle_{\mathcal{D}}}

\newcommand{\ad}{a_\mathcal{D}}
\newcommand{\da}{\dot{a}_\mathcal{D}}
\newcommand{\dda}{\ddot{a}_\mathcal{D}}

\newcommand{\Q}{\mathcal{Q}_{\mathcal{\D}}}
\newcommand{\reff}{\varrho_{\mathcal{D}}^{eff}}
\newcommand{\peff}{P_{\mathcal{D}}^{eff}}

\newcommand{\du}{\partial_{1}}

\begin{document} 

\section{Introduction}

Large-scale structures in the Universe grew by gravitational instability around primordial seed perturbations generated during inflation. 
In this context, it is assumed that the matter  mainly consists of Cold Dark Matter (CDM), which, prior to caustic formation, behaves like a perfect  fluid of dust. 
The study of the gravitational dynamics of CDM can be performed by different techniques and approximations, depending on the specific range of scales under 
investigation. 

The most widely used approximation is standard perturbation theory (see e.g. Ref.~\cite{kodama}), which is an expansion in powers 
of the amplitude of fluctuations around a Friedmann-Robertson-Walker (FRW) background.
The range of applicability of this perturbation technique is that of small fluctuations around a FRW background, with no extra limitations on scale. 
Going to higher perturbative orders generally helps to follow the gravitational instability process on a longer time-scale and to account for non-linear 
phenomena, i.e. to resolve smaller scales.

Another important approximation scheme is the Newtonian one, which is known to produce accurate results as long the gravitational field is weak and the motion of 
both sources (i.e. CDM fluid elements) and test particles is slow. In the cosmological framework, the range of validity of the Newtonian approximation 
concerns scales much larger than the Schwarzschild radius of collapsed bodies and much smaller than the Hubble horizon (see e.g. Ref.~\cite{peebles80}). 
This approximation is usually carried out in the Poisson gauge, where the FRW metric is perturbed by a single lapse function which 
represents the peculiar gravitational potential sourced by matter density fluctuations. This approach provides the well-known Newtonian Eulerian 
description of cosmological structure formation in the expanding universe. The extension of this approximation to deal with fast moving test particles (like photons)
is the so-called {\em weak-field} limit, where a scalar potential is accounted for in the spatial part of the line-element. 

In the context of {\it Newtonian gravity}, departures from the FRW background may be also analyzed in the Lagrangian frame. 
In the unperturbed situation, observers comoving with the Hubble flow are Lagrangian, but when density fluctuations are taken into account, 
they become Eulerian observers with respect to the peculiar motion. The equations describing the dynamics of CDM, which we model here through irrotational dust, 
are written in Lagrangian coordinates, in terms of the displacement vector $\mathbf{\mathcal{S}}$ from Lagrangian to Eulerian coordinates. 
As in the Eulerian case, it is impossible to work out the general analytical solution for $\mathbf{\mathcal{S}}$ and a perturbative approach is again 
introduced by means of an expansion in powers of the displacement vector, the background being once more represented by the FRW model. 
The linear result is the so-called Zel'dovich approximation, \cite{zeldovich70a}. The peculiarity of this treatment, at any order, is that, while 
the displacement vector is calculated from the equations at the required perturbative order, all the other dynamical variables, such as the mass 
density, are calculated exactly from their non-perturbative definition. Since the equations in Lagrangian coordinates are intrinsically non-linear 
in the density, what comes out is a fully non-linear description of the system, which, though not being generally exact, ``mimics'' the true non-linear 
behavior. This perturbation treatment basically exploits the advantages of the Lagrangian picture, leading, in particular, to a more accurate 
description of high density regions. Its limitations are generally set by the emerging of caustic singularities.

The Newtonian analysis can be improved by a post-Newtonian (PN) approach. In the Newtonian approximation two conditions must be satisfied: 
weak gravitational field and slow motion. Relaxing one of these two conditions leads to two different improvements of the Newtonian description. 
The PN approximation is suitable for a system of slowly moving particles bound together by gravitational forces and can be useful in cosmology 
(see Refs.~\cite{SzekeresR1, SzekeresR2} ), as it can be used to account for the moderately non-linear gravitational field generated 
during the highly non-linear stage of the evolution of matter fluctuations. 
The PN approximation in the cosmological framework has been studied in Eulerian (or more generally non-comoving) coordinates 
in Refs.~\cite{futa1,futa2,tom1,tom2,shi}, while in Ref.~\cite{CC}
a hybrid approximation scheme is proposed which upgrades the weak-field limit of Einstein's field equations to account for post-Newtonian scalar and vector metric 
perturbations and for leading-order source terms of gravitational waves, while including also the first and second-order perturbative approximations. 
An alternative approach has been recently proposed in Ref.~\cite{Baumann}, whose authors introduce an effective fluid description for small-scale non-linearities 
in the framework of General Relativity. The PN approximation to cosmological structure formation within the Lagrangian picture has 
been studied by Matarrese \& Terranova in Ref.~\cite{mater}, whose approach will be the basis for our analysis here. 

The aim of this paper is to find a PN metric describing the non-linear stages of the gravitational instability in an Einstein-de Sitter universe,  
adopting the Lagrangian picture i.e. the synchronous and comoving gauge. 
In this gauge the PN approximation is formally obtained via an expansion in inverse powers of the speed of light $c$. 
We choose the leading order solution of our expansion to be the Newtonian metric related to the Zel'dovich approximation, in the very special 
case of globally planar dynamics due to a perturbation seed $\varphi$ depending on a single Lagrangian coordinate.

Our approach here is very different from those of Refs.~\cite{kasai,russ}, who proposed relativistic generalizations of the Zel'dovich approximation.
Ref.~\cite{kasai} introduced a relativistic, tetrad-based, perturbative approach, which is then solved to linear order and used to 
obtain non-perturbative expressions for the velocity-gradient tensor and mass density. 
The solution of Ref.~\cite{russ} is instead equivalent to a relativistic second-order perturbation theory treatment in the synchronous and comoving gauge, in which 
all quantities (metric, velocity-gradient tensor and mass density) are calculated at second order, thereby partially missing
the non-perturbative character of the Zel'dovich approximation. 
Our approach instead aims at obtaining a non-perturbative description of both metric and 
fluid properties (velocity-gradient tensor and mass density), within the post-Newtonian approximation of General Relativity: our expansion in inverse powers of the 
speed of light is fully non-perturbative from the point of view of standard perturbation theory; thus our results contain all
second and higher-order terms of standard perturbation theory calculations, as long as they are post-Newtonian and one deals with the plane-parallel dynamics.
In our approximation scheme the Zel'dovich solution represents the {\em Newtonian background} over which PN corrections can 
be computed as small perturbations. 

The plan of the paper is as follows. In Section 2 we introduce the relativistic equations governing the dynamics of irrotational dust. 
Section 3 deals with the Newtonian limit. Section 4 is devoted to the PN approximation: we first describe the 
Zel'dovich Newtonian background chosen as a basis for our PN expansion, showing explicitly all the dynamical quantities involved. Then, we 
obtain and solve analytically the PN equations for the metric. In particular, we show that our solution accounts for the non-Gaussianity 
of primordial cosmological perturbations. To conclude this section, we briefly discuss the accuracy of our perturbative expansion.
Section 5 deals with an application of our solution: we use our metric in the context of the back-reaction proposal. To be more specific, 
we give an estimate of the PN contribution to the kinematical back-reaction, mean spatial curvature, average scale-factor and expansion rate, as defined in Refs.~\cite{buchertGRdust, buchertGRfluid}.
Section 6 provides a comparison between our PN solution and the Szekeres metric. Conclusions are drawn in Section 7.
 
\section{Relativistic dynamics of irrotational dust in the Lagrangian picture}

In this section we recall the relativistic equations governing the evolution of irrotational dust
following the Arnowitt-Deser-Misner (ADM) formalism, \cite{ADM}.  
Calculations are made in the synchronous and comoving gauge. In this gauge both the world-lines of the observers coincide with the geodesics of the 
matter and the proper time along the geodesics coincides with the cosmic time. This reference frame is therefore Lagrangian. Actually, the possibility of 
making these two gauge choices simultaneously is a peculiarity of irrotational dust, which holds also in the non-linear stage. \\
The line element is given by \footnote{We use Greek letters for purely spatial indices, $\a,\,\b...=1,\,2,\,3$, whereas Latin ones $a,\,b..=0,\,1,\,2,\,3$ label space-time indices.}
\begin{equation}
 ds^{2}=-c^{2}dt^{2}+h_{\alpha\beta}(t,\mathbf{q})dq^{\alpha}dq{^\beta}\,,
\end{equation}
where $q_{\alpha}$ are the spatial Lagrangian coordinates and $t$ is the proper time of our fluid elements. The matter stress-energy tensor 
reads $T^{ab} = \varrho\, c^2 u^a u^b$ with $\varrho$ the mass density and $u^a$ the fluid four-velocity (normalized to 
$u^a u_a = -1$), with components $u^a = (1,0,0,0)$. The ADM equations can be written in terms of the velocity-gradient tensor defined as
\begin{equation}
 \t^{\a}_{\b}=c u^{\a}_{~;\b} =\frac{1}{2}h^{\a\ss}\dot{h}_{\ss\b}\,,
\end{equation} 
where dots denote partial differentiation with respect to the time $t$ and semi-colons stand for covariant differentiation.
The tensor $\t^\a_{\b}$ represents the extrinsic curvature of constant $t$ hyper-surfaces. 

The 10 Einstein equations split into 4 constraints and 6 evolution equations. 
The time-time component of the Einstein equations is the so-called energy constraint of the ADM approach, which reads 
\begin{equation} \label{energyh}
\Theta^{2}-\Theta^{\alpha}_{\beta}\Theta^{\beta}_{\alpha}+c^{2} \,^{(3)}\!R=16\pi G\varrho\,, 
\end{equation}
where the volume-expansion scalar $\Theta$ is the trace of the velocity-gradient tensor and $\,^{(3)}\!R$ is the trace of the 3-dimensional Ricci 
tensor $\,^{(3)}\!R^\alpha_\beta$ of the spatial hyper-surfaces of constant time. 

The space-time components give the momentum constraint,
\begin{equation}
 \Theta^{\alpha}_{\beta;\alpha} = \Theta_{,\beta} \; .
\end{equation} 
The space-space components govern the evolution of the velocity-gradient tensor and represent the only truly evolution equations. They read
\begin{equation}
 \dot{\Theta}^{\alpha}_{\beta}+\Theta \Theta^{\alpha}_{\beta}+c^{2} \,^{(3)}\!R^\alpha_\beta = 4\pi G\varrho \delta^{\alpha}_{\beta}\,.
\end{equation}
Taking the trace of the latter equation and combining it with the energy constraint yields the Raychaudhuri equation for the {\em volume-expansion} scalar $\Theta$, 
\begin{equation}\label{raychaudh}
 \dot{\t}+2\Sigma^{2}+\frac{1}{3}\t^{2}+4 \pi G \r=0\,,
\end{equation}
where $\Sigma^\a_\b=\Theta^\a_\b-(1/3)\Theta\d^\a_\b$ is the {\em shear} tensor and $\Sigma^2=(1/2)\Sigma^\a_\b\Sigma^\b_\a$ its magnitude.
Finally, the continuity equation for the stress-energy tensor, $T_{\a\b}=\r u_{\a}u_{\b}$, leads to
\begin{equation}\label{masscons}
 \dot{\varrho} = - \Theta \varrho \,,
\end{equation}
whose solution reads
\begin{equation} \label{contrho}
 \r(\mathbf{q},t)=\r_{\rm{in}}(\mathbf{q})\sqrt{\frac{h_{\rm{in}}(\mathbf{q})}{h(\mathbf{q},t)}}\,,
\end{equation}
$h$ being the determinant of the spatial metric. Here and in what follows quantities with a subscript ``$\rm{in}$" 
are meant to be evaluated at some initial time $t_{\rm{in}}$.

\subsection{Conformal rescaling and FRW background subtraction}

With the purpose of studying gravitational instability in the Einstein-de Sitter background, we first factor out the homogeneous and isotropic expansion of the Universe. 
To this aim, we perform a conformal rescaling of the metric with the conformal factor $a(\tau)\propto\tau^2$, the scale factor of the Einstein de-Sitter model. 
This procedure involves the transformation of the proper time $t$ to conformal time $\tau$ via $d\tau=dt/ a(t)$.
The line-element is written in the form
\begin{equation}
 ds^{2}=a^{2}(\tau)\left[-c^{2}d\tau^{2}+\gg_{\alpha\beta}(\tau,\mathbf{q})dq^{\alpha}dq{^\beta}\right],
\end{equation} 
where $\gg_{\alpha\beta}(\tau,\mathbf{q})=h_{\alpha\beta}\left(t\left(\tau\right),\mathbf{q}\right)/a^{2}\left(t\left(\tau\right)\right)$ is the conformal spatial metric.
In order to factor out the Einstein-de Sitter background, the isotropic Hubble flow, with velocity-gradient tensor $(a'/a)\d^{\a}_{\b}$, is subtracted from the velocity-gradient 
tensor, leading to a tensor which describes gradients of the peculiar velocity only:
\begin{equation} \label{rescdec}
\vartheta^{\a}_{\b} = ac \tilde{u}^\a_{;\b} - \frac{a'} {a}\d^{\a}_{\b}=\frac{1}{2}\gg^{\a\ss}\gg'_{\ss\b}\,,
\end{equation}
where $\tilde{u}^\a = (1/a,0,0,0)$ and primes denote partial differentiation with respect to $\tau$. The tensor $\vartheta^\a_{\b}$ represents the extrinsic curvature of constant $\tau$ 
spatial hypersurfaces. 
The matter content is also written in terms of the density contrast, defined as the dimensionless deviation of the matter density from that of the Einstein-de Sitter background, 
$\r_b(\tau)=3/(2\pi G a^2\tau^2)$,
\begin{equation}
 \d(\tau,\mathbf{q}):=\frac{\r(\tau,\mathbf{q})-\r_{b}(\tau)} {\r_b(\tau)}\,.
\end{equation} 
The ADM equations are then recast in a more convenient form describing the evolution of the peculiar velocity-gradient tensor. The energy and momentum constraints become
\begin{eqnarray}
  \vartheta^{2}-\vartheta^{\alpha}_{\beta}\vartheta^{\beta}_{\alpha}+\frac{8} {\tau}\vartheta+c^{2}\mathcal{R}&=&\frac{24}{\tau^2}\d \label{Emater} \\
\mathcal{D}_{\alpha}\vartheta^{\alpha}_{\beta}&=& \vartheta_{,\b} \,, \label{mommater}
\end{eqnarray}
where $\mathcal{D}_{\alpha}$ is the covariant derivative associated with $\gg_{\a\b}$ and $\mathcal{R}^\a_\b=a^2 \,^{(3)}\!R^\a_\b$ ($\mathcal{R}=a^2 \,^{(3)}\!R$) is the conformal three-dimensional (scalar) curvature. 
The evolution equation becomes
\begin{equation} 
 \vartheta^{\a'}_{\b}+\frac{4} {\tau}\vartheta^{\a}_{\b}+\vartheta\vartheta^{\a}_{\b}+\frac{2}{\tau}\vartheta\d^{\a}_{\b}+c^{2}\mathcal{R}^{\a}_{\b}=\frac{6}{\tau^2}\d\d^{\a}_{\b}\,.
\end{equation} 
After replacing the density from the energy constraint, this equation is written as
\begin{eqnarray} \label{evolmater}
 0&=&\vartheta^{\a'}_{\b}+\frac{4 }{\tau}\vartheta^{\a}_{\b}+\vartheta\vartheta^{\a}_{\b}+\frac{1}{4}\left(\vartheta^{\mu}_{\nu}\vartheta^{\nu}_{\mu}- \vartheta^{2}\right)\d^{\a}_{\b}+ \nonumber \\
&& +\frac{c^{2}}{4}\left[4\mathcal{R}^{\a}_{\b}-\mathcal{R}\d^{\a}_{\b}\right].
\end{eqnarray}
The Raychaudhuri equation describes the evolution of the peculiar volume expansion scalar and reads
\begin{equation}
 \vartheta'+\frac{2}{\tau}\vartheta+\vartheta^{\mu}_{\nu}\vartheta^{\nu}_{\mu}+\frac{6}{\tau^2}\d=0\,.
\end{equation} 
Finally, the solution of the continuity equation~(\ref{contrho}) is written in terms of the contrast $\d$ as
\begin{equation} \label{dmater}
 \d(\mathbf{q},t)=\left(1+\d_{\rm{in}}(\mathbf{q})\right)\sqrt{\frac{\gg_{\rm{in}}(\mathbf{q})}{\gg(\mathbf{q},t)}} -1\,,
\end{equation}
where $\gg$ is the determinant of the conformal spatial metric $\gg_{\a\b}$. 

The main advantage of this formalism is that there is only one dimensionless variable
in the equations, namely the spatial metric tensor $\gg_{\a\b}$, which is present with its partial time
derivatives through $\vartheta^{\a}_{\b}$ and with its spatial gradients through the spatial Ricci tensor $\mathcal{R}^{\a}_{\b}$. A relevant advantage of having a single 
tensorial variable, for the following PN expansion,
is that there can be no extra powers of $c$ hidden in the definition of different quantities.

\section{Newtonian approximation}

The Newtonian equations in Lagrangian form can be obtained from the full relativistic equations of the previous section by an expansion in inverse 
powers of the speed of light. As a consequence of our gauge choice, however, no odd powers of $c$ appear in the equations, so the expansion 
parameter is $1/c^{2}$. The spatial metric is then expanded in the form
\begin{equation}
 \gg_{\a\b}=\overline{\gg}_{\a\b}+\frac{1}{c^{2}} w_{\a\b}+\mathcal{O}\left(\frac{1}{c^4}\right).
\end{equation} 

Let us first concentrate on the Newtonian metric $\overline{\gg}_{\a\b}$. 
The Newtonian limit ($c\rightarrow\infty$) of the energy constraint and evolution equation requires that the spatial Ricci tensor is zero. Thus, the Newtonian 
result of vanishing spatial curvature is recovered (see Refs.~\cite{ellis71, mater}). This important conclusion implies that $\overline{\gg}_{\a\b}$ can be transformed 
to the Euclidean metric $\d_{\a\b}$ globally. In other words, at each time $\tau$ there exist global Eulerian observers comoving with the Hubble flow for which 
the components of the metric are $\d_{\a\b}$. This means that, according to the tensor transformation law, we can write
\begin{equation} \label{globalflat}
 \overline{\gg}_{\a\b}=\d_{\mu\nu}\mathcal{J}^{\mu}_{\a}\mathcal{J}^{\nu}_{\b}\,,
\end{equation}  
where 
\begin{equation} \label{jaco}
\mathcal{J}^{\mu}_{\a}=\d^{\mu}_{\a}+\frac{\partial\mathbf{\mathcal{S}}^{\mu}}{\partial q^{\a}} 
\end{equation} 
is the Jacobian of the transformation
\begin{equation} \label{eulcomcomp}
\mathbf{x}(\mathbf{q},\tau)=\mathbf{q}+\mathbf{\mathcal{S}}(\mathbf{q},\tau)
\end{equation} 
and $x^\mu$, $(\mu=1, 2, 3)$, are the Eulerian (comoving with the isotropic FRW background) coordinates. The tensor $\partial\mathcal{S}^{\mu}/\partial q^{\a}$ is called ``deformation tensor''.
All the information about the motion is contained in the mapping between the Eulerian and Lagrangian coordinates, whose evolution is described by 
the Raychaudhuri equation and the momentum constraint. 
In fact, contrary to the evolution equation and the energy constraint, the Raychaudhuri equation and the momentum constraint contain no explicit powers of $c$, 
thus preserve their form in going to the Newtonian limit. These equations determine the background Newtonian metric $\overline{\gg}_{\a\b}$: they are indeed 
the Newtonian equations in the Lagrangian approach to irrotational dust dynamics (see Ref.~\cite{ellis71}).
Thus, it makes sense to recast these equations in terms of the map between comoving Eulerian and Lagrangian coordinates. One can equivalently write the equations 
in terms of either the displacement vector, as in Refs.~\cite{buchert92, catelan95}, and deformation tensor or in terms of the Jacobian matrix of the map, as in Ref.~\cite{mater}. Here, following Ref.~\cite{mater}, 
we consider the Newtonian equations in terms of the Jacobian matrix~(\ref{jaco}). They read
\begin{eqnarray} 
 \mathcal{J}^{\a}_{\mu}\mathcal{J}^{\mu''}_{\a}+\frac{2}{\tau}\frac{\mathcal{J}'}{\mathcal{J}}&=&\frac{6}{\tau^2}\left(1-\frac{1}{\mathcal{J}}\right) \\
 \varepsilon^{\a\b\gg}\mathcal{J}^{\mu}_{\b}\mathcal{J}'_{\mu\gg}&=&0\,,
\end{eqnarray}
where $\mathcal{J}$ is the determinant of the Jacobian matrix. We assumed for simplicity $\d_{\rm{in}}=0$ and used the residual gauge freedom of the 
synchronous and comoving gauge to set $\mathcal{J}_{\rm{in}}=1$ in the Newtonian limit, as in Ref.~\cite{mater}.

It is impossible to follow the dynamics in an exact analytical way and one is forced to use approximation techniques to solve these equations. Here the 
perturbative expansion is performed in terms of the Jacobian matrix. The key point is that a slight perturbation of the Lagrangian particle paths carries 
information on the non-linear dynamics, as may be seen directly from the expression of the density contrast
\begin{equation} \label{densjaco}
 \d=\frac{1}{\mathcal{J}}-1\,,
\end{equation}
which shows that, even if the displacement vector is small, the corresponding density contrast could be large.
In other words, already the first-order Lagrangian equations describe the mildly non-linear regime of the gravitational instability. 

Expanding and solving the previous equations to first order, the well-known Zel'dovich approximation is recovered:
\begin{equation}
\mathbf{x}(\mathbf{q},\tau)=\mathbf{q}+D(\tau)\nabla \Phi (\mathbf{q})\,,
\end{equation} 
where $D(\tau)\propto a(\tau)\propto \tau^{2}$ is the growing mode solution for the Einstein-de Sitter model and $\Phi(\mathbf{q})$ has to be ascribed to the chosen initial conditions.

\section{Post-Newtonian approximation}

\subsection{Characterization of the Newtonian background}

The starting point of our PN expansion is the Newtonian background described by the Zel'dovich approximation, with the peculiar gravitational potential 
depending on the conformal time and on the Lagrangian coordinate $q_1$ only. As it is well known, in the particular case of planar perturbations the 
Zel'dovich approximation yields an exact solution of the Newtonian equations.
The {\it Zel'dovich solution} is given by 
\begin{eqnarray}
x_1 &=&q_1 + D(\tau)\du\Phi_{\rm{in}}(q_1) \nonumber \\
x_2 &=&q_2 \nonumber \\
x_3 &=&q_3\,.
\end{eqnarray}
where $\du=\partial/\partial q_1$ and the potential $\Phi_{\rm{in}}$ is defined so that $\du^{2}\Phi_{\rm{in}}=-\d_{\rm{in}}/D_{\rm{in}}$ and is related to the initial peculiar 
gravitational potential $\varphi$ by the cosmological Poisson equation, yielding
\begin{equation}
 \Phi_{\rm{in}}=-\frac{\varphi}{4\pi G a_{\rm{in}}^{2}\r_{b,\rm{in}}}\,.
\end{equation} 
The Jacobian matrix is 
\begin{equation} \label{caustjaco}
 \mathcal{J}^{\a}_{\b} = 
\pmatrix{
 1-\tau^2\du^2\varphi/6 &  &  \cr
 & 1 &  \cr
 &  & 1}\,.
\end{equation} 
For the following calculations, it is useful to define the function $f$
\begin{equation}
 f := D(\tau)\du^{2} \Phi_{\rm{in}}=-\frac{\tau^2}{6}\du^{2}\varphi
\end{equation} 
and the function $\eta \equiv \ln \left(1+f\right)$.
Hereafter, the peculiar gravitational potential is meant to be evaluated at the initial time $\tau_{\rm{in}}$ and the subscript ``$\rm{in}$" is dropped for notational convenience.

The components of the metric in Lagrangian coordinates are given by (recall Eq.~(\ref{globalflat}))
\begin{equation} \label{ggzel}
 \overline{\gg}_{\alpha\beta}=\d_{\ss\omega}\left[\d^{\ss}_{\a}+D(\tau)\partial^{\ss}\partial_{\a}\Phi\right]\left[\d^{\omega}_{\b}+D(\tau)\partial^{\omega}\partial_{\b}\Phi\right],
\end{equation}
where we used the fact that at first order in the displacement vector covariant and partial derivatives with respect to the coordinates $q_{\a}$ coincide, since the Christoffel 
symbols are second-order quantities. 
In our case $\Phi_{\rm{in}}=\Phi_{\rm{in}}(q_1)$ and for the Zel'dovich metric we find
\begin{equation} \label{metricor}
 \overline{\gg}_{\alpha\beta} = 
\pmatrix{
 (1-\tau^{2}\du^{2}\varphi/6)^{2}&  &  \cr
 & 1 &  \cr
 &  & 1}
\end{equation} 
or in more compact form
\begin{equation}
 \overline{\gg}_{\alpha\beta} = 
\pmatrix{
 (1+f)^{2}&  &  \cr
 & 1 &  \cr
 &  & 1}\,.
\end{equation} 

It is important to keep in mind that this Newtonian metric is non-linear with respect to the peculiar gravitational potential, thus it characterizes the mildly 
non-linear stage of the gravitational instability. 
Starting from this metric at first order in the displacement vector, all the other dynamical variables are calculated exactly.
The only non-vanishing component of the peculiar velocity-gradient tensor is $\overline{\vartheta}^{1}_{1} = \eta'$ and for the shear tensor we have 
$\overline{\ss}^{1}_{1} = 2\eta'/ 3$ and $\overline{\ss}^{2}_{2} = \overline{\ss}^{3}_{3} = -\overline{\ss}^{1}_{1}/2$. \\
Finally the density contrast~(\ref{densjaco}) takes the form
\begin{equation}\label{caustden}
\overline{\d}=\frac{1}{1+f}-1\,.
\end{equation}

\subsection{Post-Newtonian expansion} \label{PNsection}

For the PN expansion, we write the metric in the form
\begin{equation}
 \gg_{\a\b}=\overline{\gg}_{\a\b}+\frac{1}{c^{2}} w_{\a\b}\,.
\end{equation} 
For consistency with our Newtonian background solution, in which the peculiar gravitational potential 
depends only on $q_1$, the PN perturbation $w_{\a\b}$ can be assumed to depend on the conformal time and on the Lagrangian coordinate $q_1$ only.

A clue for the form of the perturbation $w_{\a\b}$ follows from the initial conditions of cosmological perturbations. Even though our case of globally planar
dynamics is purely a toy-model we prefer to assume that our initial perturbation seed is consistent with having been
generated during inflation in the early Universe, so that our analytical results can give us a hint of what happens in the real Universe, where the initial perturbation seed 
is a random field that depends on all the spatial coordinates.  
We then set our initial conditions at the end of inflation, effectively coinciding with $\tau_{\rm{in}}=0$. Considering only scalar perturbations from the 
Einstein-de Sitter Universe, we have at early times 
\begin{equation} \label{ci}
 \gg_{\a\b}=\left(1-\frac{10}{3c^2}\varphi\right)\d_{\a\b}-\frac{\tau^2}{3}\partial_\a\partial_\b\varphi \,.
\end{equation}  
We can use the residual gauge freedom of the synchronous and comoving gauge to set the Newtonian perturbation to zero (see Ref.~\cite{mater}). The initial metric 
perturbation is therefore given by a diagonal PN part. Thus, for the PN perturbation at initial time we have
\begin{equation}
 w^\a_{\b \rm{in}} =-\frac{10}{3}\varphi \delta^\a_\b \;.
\end{equation}  

Starting from these initial conditions, we can assume that the evolution does not switch on the off-diagonal components of the PN metric, i.e. that $w^{\a}_{\b}$ 
with $\a\neq\b$ vanish at any time. This assumption derives from the physical picture of our one-dimensional dynamics. The mass 
distribution whose self-gravity generates a one-dimensional potential is made of parallel sheets of matter. For every point $\mathbf{q}$ the peculiar 
velocity has the same direction as the spatial derivative of the peculiar potential, thus the matter moves only in the direction perpendicular to the sheets. 
The collapse of this structure cannot involve tensor perturbations, which would lead to the emission of gravitational waves, because it cannot undergo 
any alteration of its shape. However, there is surely a scalar trace part in the PN metric, arising from our inflationary initial conditions. 
In addition, because of the asymmetry in the $q_1$ spatial direction, the function $w^{1}_{1}$ is assumed to differ from $w^{2}_{2}$ and $w^{3}_{3}$, 
whereas the latter functions can only be equal.

Therefore, the PN expansion is performed according to the following ansatz for the metric\footnote{The indices of the perturbation are lowered (raised) 
with the background metric $\overline{\gg}_{\a\b}$ ($\overline{\gg}^{\a\b}$).}
\begin{eqnarray} \label{ansatz}
  \gamma_{11}&=& (1+f)^{2}+\frac{1}{c^{2}}(1+f)^{2}g \\
 \gamma_{22}&=& 1+\frac{1}{c^{2}} h\\
  \gamma_{33}&= &1+\frac{1}{c^{2}} h \,,
\end{eqnarray} 
with initial conditions
\begin{equation} \label{cig}
 g_{\rm{in}}=h_{\rm{in}}=-\frac{10}{3}\varphi\,.
\end{equation} 

The PN expansion of the momentum constraint and of the Raychaudhuri equation gives
\begin{eqnarray}
\eta^\prime \du h&=&(\du h)^\prime \\
g''+2h''+\frac{2}{\tau}(g'+2h')+2\eta'g'&=&-\frac{6}{\tau^{2}}\frac{\left(g_{\rm{in}}+2h_{\rm{in}}\right)-(g+2h)}{1+f}\,.
\end{eqnarray} 
We also have the energy constraint and evolution equation, connecting the PN scalar curvature with Newtonian kinematical quantities:
\begin{eqnarray} \label{encheck}
\frac{8}{\tau}\eta'+\frac{1}{(1+f)^{2}}\left(-2\du^{2} h+2\du\eta\du h\right)&=&\frac{24}{\tau^{2}}\left(\frac{1}{1+f}-1\right) \\
 \label{evcheck}
 \eta''+\frac{4}{\tau}\eta'+\frac{1}{4}\frac{1}{(1+f)^{2}} \left(-2\du^{2} h+2\du\eta\du h\right)&=&0\,.
\end{eqnarray} 

The momentum constraint is an equation for the spatial derivative of the function $h$. Setting $m \equiv \du h$, it reads
\begin{equation}
 \eta'm=m'\,,
\end{equation} 
with initial condition $m_{\rm{in}}=(-10/3)\du \varphi$. The solution reads
\begin{equation}
m=-\frac{10}{3}\du \varphi\left(1-\frac{\tau^{2}}{6}\du^{2}\varphi\right).
\end{equation}
Then, by spatial integration we obtain 
\begin{equation} \label{lamiah}
 h= -\frac{10}{3} \varphi+\frac{5}{18}\tau^{2}(\du\varphi)^{2}+ C_0(\tau)\,,
\end{equation} 
where the homogeneous mode $C_0(\tau)$ is a time-dependent constant of integration (w.r.t. $q_1$).
We can use the Newtonian evolution equation and energy constraint to check the consistency of this solution: 
substitutions of~(\ref{lamiah}) in~(\ref{encheck}) and~(\ref{evcheck}) leads to the identity.

Note that the initial condition for the function $h$ sets $C_0 (\tau_{\rm{in}}) = 0$. In addition, the function $C_0(\tau)$ is an additive 
term in the perturbation $h$ which would modify the background dynamics even in the absence of any initial perturbation (i.e. for $\varphi=0$).
Therefore, for consistency, we set $C_0(\tau)=0$ for all times.

The Raychaudhuri equation becomes an equation for the function $g$ only, whose solution reads
\begin{equation} \label{solg}
 g= -\frac{10}{3} \varphi+\frac{\tau^2 C_2}{5\left(-6+\tau^2\du^{2}\varphi\right)}+\frac{ C_1}{t^3\left(-6+\tau^2\du^{2}\varphi\right)}-\frac{5 
 \tau^4\du^{2}\varphi(\du\varphi)^{2}}{21\left(-6+\tau^2\du^{2}\varphi\right)}\,,
\end{equation} 
where $C_1$ and $C_2$ are integration constants.
Consistency with our initial conditions, Eq.~ (\ref{cig}), requires $C_1=0$.

In conclusion, the PN metric reads
\begin{eqnarray} \label{soluz}
\gamma_{11}&=& \left(1-\frac{\tau^{2}}{6}\du^{2}\varphi\right)^{2}+\frac{1}{c^{2}} \left(-6+\tau^{2}\du^{2}\varphi\right)\left(\frac{21
 \tau^2C_2-25\tau^4\du^{2}\varphi(\du\varphi)^{2}-350\varphi \left(-6+\tau^2\du^{2}\varphi\right)}{3780}\right) \nonumber  \\
 \gamma_{22}&=& 1+\frac{1}{c^{2}} \left(-\frac{10}{3} \varphi+\frac{5}{18}\tau^{2}(\du\varphi)^{2}\right) \nonumber\\
 \gamma_{33}&=& 1+\frac{1}{c^{2}} \left(-\frac{10}{3} \varphi+\frac{5}{18}\tau^{2}(\du\varphi)^{2}\right).
\end{eqnarray}

\subsubsection{Determination of the integration constant $C_2$}

In the metric~(\ref{soluz}) the initial condition $C_{2}(q_1)$ is still undetermined. To determine it we take advantage of the results 
obtained in Refs.~\cite{BMR, nongaus}. The authors consider the primordial non-Gaussianity set at inflationary epochs on super-Hubble 
scales. At later times, cosmological perturbations re-enter the Hubble radius. They show how the information on the primordial 
non-Gaussianity, set on super-Hubble scales, flows into smaller scale using a general relativistic computation. Their calculations, which
are performed in the synchronous and comoving gauge, show how the primordial non-Gaussianity affects the PN part of the density contrast at second order.  
Once again the use of inflationary initial conditions in our case of globally planar dynamics is justified by our ultimate goal of having a hint on what happens 
in the fully three-dimensional dynamics.   

First of all, we consider our fully non-linear PN expression for the density contrast. The PN contribution is given by $\d^{PN}=\left(1/2\right) 
\left(1+\overline{\d}\right)\left(w_{\rm{in}}-w\right)$, where $w$ is the trace of the PN perturbation $w^\a_\b$ of the metric 
\begin{equation}
 w=\frac{-3150\varphi\left(\tau^{2}\du^{2}\varphi-6\right)+\tau^2\left[63 C_2+50(\du\varphi)^{2}\left(2\tau^{2}\du^{2}\varphi-21\right)\right]}{315\left(\tau^{2}\du^{2}\varphi-6\right)}\,.
\end{equation} 
For a comparison with the result of Ref.~\cite{nongaus}, our expression for the density contrast
\begin{equation} \label{densPN}
   \d =\frac{\tau^{2}\du^{2}\varphi}{6-\tau^{2}\du^{2}\varphi}+\frac{1}{c^{2}}\left(\frac{21 C_2 \tau^2-
   25\tau^4\du^{2}\varphi(\du\varphi)^{2}}{35\left(6-\tau^{2}\du^{2}\varphi\right)^2}-\frac{5\tau^2(\du\varphi)^{2}}{3\left(6-\tau^{2}\du^{2}\varphi\right)}\right)
\end{equation} 
must be expanded up to second order with respect to the peculiar gravitational potential.
As usual, we split the density contrast into a first and second order part $\d=\d^{(1)}+(1/2)\d^{(2)}$, finding 
\begin{equation}
 \d=\frac{1}{6}\tau^{2}\du^{2}\varphi+\frac{C_2}{60c^{2}}\tau^{2}+\frac{C_2}{180c^{2}}\tau^{4}\du^{2}\varphi-\frac{5}{18c^{2}}\tau^{2}(\du\varphi)^{2}
 +\frac{C_2}{720c^{2}}\tau^{6}(\du^{2}\varphi)^{2}+\frac{1}{36}\tau^{4}(\du^{2}\varphi)^{2}\,.
\end{equation} 
Actually, in this expression one can only be sure about the order of the terms that do not contain $C_2$, since the latter implicitly depends on the initial peculiar 
gravitational potential, as it will be shown. 
The first-order term, i.e. $\tau^{2}\du^{2}\varphi/6$, obviously coincides with the result of linear perturbation theory in the synchronous and 
comoving gauge and it is a Newtonian term, as it is well known.
The remaining terms are at least of second order. They read
\begin{equation} \label{lamia}
\d^{(2)}= \frac{C_2}{30c^{2}}\tau^{2}+\frac{C_2}{90c^{2}}\tau^{4}\du^{2}\varphi-\frac{5}{9c^{2}}\tau^{2}(\du\varphi)^{2}+
\frac{C_2}{360c^{2}}\tau^{6}(\du^{2}\varphi)^{2}+\frac{1}{18}\tau^{4}(\du^{2}\varphi)^{2}\,.
\end{equation} 
This expression can be compared with Eq.~(45) of Ref.~\cite{nongaus}, by specializing the latter to an Einstein-de Sitter background model and to globally 
planar perturbations\footnote{For the  Einstein-de Sitter background in (45) of Ref.~\cite{nongaus} we set: $\Omega_{0m}=1$, $f\left(\Omega_{0m}\right)=
1$, $\mathcal{H}_{0}=2/\tau_{0}$, where the subscript "$0$" denotes the present time, and $D_{+}(\tau) =\tau^{2}/\tau_0^{2}$ is the linear growing mode solution.}. 
It reads (in $c=1$ units) 
\begin{equation} \label{la45}
 \d^{(2)}=\frac{10}{9}\left(\frac{3}{4}-a_{NL}\right)\tau^{2}(\du\varphi)^{2}+\frac{10}{9}\left(2-a_{NL}\right)\tau^{2}\varphi(\du^{2}\varphi)+\frac{1}{18}\tau^{4}(\du^{2}\varphi)^{2}\,,
\end{equation} 
where the deviation of the parameter $a_{NL}$ from unity measures the strength of the initial (i.e. inflationary) non-Gaussianity (see Ref.~\cite{nongaus} for more details). 
Looking at these expressions, we first note that the second-order Newtonian term, i.e. $(1/18)\tau^{4}(\du^{2}\varphi)^{2}$, is the same in 
both expression, as it should be. 
For what concerns the form of $C_2(q_1)$, it should be recalled that it is the initial condition of the PN growing mode $\propto\tau^{2}$ 
in the solution~(\ref{solg}). Thus, we already know that it must be (at least) a second-order term, since the analogous first-order term 
is Newtonian. 
The next step is to recognize the PN terms in Eq.~(\ref{la45}). Although the explicit powers of $c$ are not shown, one knows from dimensional 
analysis that the second order, i.e. $\propto\varphi^{2}$, PN terms should be $\propto\tau^{2}$ and contain two spatial derivatives, or they 
should be $\propto\tau^{4}$ with four spatial derivatives and so on, in order to have the correct powers of $c$ and to be dimensionless, 
second-order quantities. The PN terms in Eq.~(\ref{la45}) are then 
\begin{equation}
 \frac{10}{9}\left(\frac{3}{4}-a_{NL}\right)\tau^{2}(\du\varphi)^{2} 
\end{equation} 
and 
\begin{equation}
 \frac{10}{9}\left(2-a_{NL}\right)\tau^{2}\varphi(\du^{2}\varphi)\,.
\end{equation} 

Notice that it is precisely the PN terms which bring all the relevant information about (quadratic) primordial non-Gaussianity.
Note also that PN terms $\propto\tau^{4}$ with four spatial derivatives and PPN terms are absent (indeed, they would appear at third order in perturbation theory).
Now, in the expression~(\ref{lamia}), it is explicitly shown that $\left(C_2/30c^{2}\right)\tau^{2}$ is a PN term. Therefore, $C_2$ must contain two 
spatial derivatives. This fact completely determines the form of  $C_2$: the most general expression that can be constructed is 
\begin{equation} \label{c2}
 C_2=A(\du\varphi)^{2}+B\varphi(\du^{2}\varphi)\,.
\end{equation} 
At this point, notice that the PN terms in Eq.~(\ref{lamia}) 
\begin{equation}
 \frac{C_2}{90c^{2}}\tau^{4}\du^{2}\varphi 
\end{equation} 
and
\begin{equation}
 \frac{C_2}{360c^{2}}\tau^{6}(\du^{2}\varphi)^{2}
\end{equation} 
are actually third and fourth-order terms, respectively. Substitution of Eq.~(\ref{c2}) in Eq.~(\ref{lamia}) leads to
\begin{equation}
  \left(\frac{A}{30c^{2}}-\frac{5}{9c^{2}}\right)\tau^{2}(\du\varphi)^{2}+\frac{B}{30c^{2}}\tau^{2}\varphi(\du^{2}\varphi)
\end{equation}
for our second-order PN terms.
By comparison with Eq.~(\ref{la45}), we find $A$ and $B$ in terms of $a_{NL}$ we finally obtain
\begin{equation}
C_2=\frac{25}{3}\left[\left(1-4\left(a_{NL}-1\right)\right)(\du\varphi)^{2}+\left(4-4\left(a_{NL}-1\right)\right)\varphi(\du^{2}\varphi)\right]. 
\end{equation}

The final expression of our PN metric reads
 \begin{eqnarray} \label{soluzione}
\gamma_{11}&= &\left(1-\frac{\tau^{2}}{6}\du^{2}\varphi\right)^{2}+\frac{1}{c^{2}}\left\{ \left[\frac{5}{108}
\tau^2\left(\left(4\left(a_{NL}-1\right)-1\right)\left(\du\varphi\right)^2+\left(4\left(a_{NL}-1\right)-4\right)
\varphi\du^{2}\varphi\right)+ \right. \right. \nonumber \\
&& \hphantom{\times\biggl\{}\left. \left. +\frac{5}{576}\tau^4\du^{2}\varphi\left(\du\varphi\right)^2 
\right]\left(6-\tau^2\du^{2}\varphi\right)-\frac{5}{54}\varphi\left(6-\tau^2\du^{2}\varphi\right)^2\right\} \nonumber \\
\gamma_{22}&=& 1+\frac{1}{c^{2}} \left(-\frac{10}{3} \varphi+\frac{5}{18}\tau^{2}(\du\varphi)^{2}\right) \\
\gamma_{33}&=& 1+\frac{1}{c^{2}} \left(-\frac{10}{3} \varphi+\frac{5}{18}\tau^{2}(\du\varphi)^{2}\right). \nonumber
\end{eqnarray} 

These expressions for the metric represent the main result of this paper:  they provide
the post-Newtonian extension of the well-known Zel'dovich solution for plane-parallel cosmological dynamics in Newtonian gravity.

\subsubsection{Convergence of the perturbation series}

The actual convergence of the perturbative series requires that the PN metric is much smaller than the background 
Newtonian one. To estimate the order of magnitude of the different contributions, one should
keep in mind that, on sub-Hubble scales, the peculiar gravitational potential is suppressed with respect to the matter density contrast by the 
square of the ratio of the proper scale of the perturbation $\lambda_{proper}$ to the Hubble radius $r_H=cH^{-1}$.
Indeed, from the cosmological Poisson equation, 
\begin{equation}
 \frac{\varphi}{c^{2}} \sim \left( \frac{\lambda_{proper}}{cH^{-1}}\right)^{2}\d\,,
\end{equation} 
which makes it clear that the gravitational potential divided by the square of the speed of light can remain small even on 
scales characterized by a large density contrast (only provided $\vert \d \vert \ll (c H^{-1}/\lambda_{\rm proper})^2$),  which is indeed 
at the basis of the well-known validity of the Newtonian approach to cosmological structure formation.
We must also recall that, in the Newtonian limit, the square of the peculiar velocity is of the same order as the peculiar gravitational potential and both remain small even in the non-linear regime of structure formation.  \\
Let us then consider the various terms in the metric of Eq.~(\ref{soluz}). For $\gg_{11}$ we find
\begin{equation}
\mathcal{O}\left(\frac{1}{\left(1+\overline{\d}\right)^2}\right)+\mathcal{O}\left(\frac{\varphi/c^2}{1+\overline{\d}}\right)+\mathcal{O}\left(\frac{\varphi/c^2}{\left(1+\overline{\d}\right)^2}\right),
 \end{equation}
where the first term belongs to the Newtonian part. Similarly, for $\gg_{22}=\gg_{33}$ we have
\begin{equation}
 \mathcal{O}(1)+\mathcal{O}\left(\frac{\varphi}{c^{2}}\right).
\end{equation} 
It is clear that the PN terms are sub-leading.

\section{A post-Newtonian estimation of kinematical back-reaction}

It has been proposed that the observed increase in the expansion rate of the Universe could be due to the back-reaction of the 
of non-linear sub-horizon cosmic structures on the background Universe expansion \cite{ras,KMR,notari}. The issue of 
the quantitative relevance of back-reaction effects is a largely controversial one 
(see, e.g. Ref.~\cite{buchreview} for a review). 
It is well-known that both standard perturbative treatments even at higher than linear order (see e.g. the discussion in Ref.~\cite{KMR}) and the 
Newtonian approximation (see in this respect the analysis 
of Ref.~\cite{buchertNEWTdust}) are totally inadequate to correctly evaluate the relevance of back-reaction terms in the average Einstein's equations 
(see also Ref.~\cite{rasnewt}). 
For these reasons it is tempting to evaluate how PN terms can affect back-reaction.  
We will then examine this issue here using the PN expression of the metric just obtained. The importance of using the general relativistic Lagrangian description 
of cosmological perturbations in connection with back-reaction has also been recently stressed in Ref.~\cite{buchlagrang}.

A set of effective Friedmann equations that describe the average dynamics of the real inhomogeneous Universe containing irrotational 
dust have been obtained by Buchert in Ref.~\cite{buchertGRdust} by smoothing the Einstein equations by a spatial average for a scalar $\Psi$ defined as
\begin{equation}
 \av{\Psi}=\frac{1}{\mathcal{V}_{\mathcal{D}}}\int_{\mathcal{D}}{\Psi\sqrt{h}~d^{3}q}\,,
\end{equation} 
where $h$ is the determinant of the metric $h_{\a\b}$ and $\mathcal{V}_\D$ is the volume of the coarse-graining comoving domain $\D$
\begin{equation}
 \mathcal{V}_{\D}=\int_{\mathcal{D}}{\sqrt{h}~d^{3}q}\,.
\end{equation}
By smoothing the scalar Einstein equations, Eqs.~(\ref{energyh}) and~(\ref{raychaudh}), the following effective Friedmann equations for 
the average scale factor $a_\D=\left(\mathcal{V}_\D/\mathcal{V}_{\D_0}\right)^{1/3}$ are obtained 
\begin{eqnarray}
 \left(\frac{\da}{a_{\D}}\right)^{2}&=&\frac{8}{3} \pi G \r_{\D}^{eff}\\
\left(\frac{\dda}{a_{\D}}\right) &=& -\frac{4}{3}\pi G\left(\r_{\D}^{eff} + \frac{3P_{\D}^{eff}}{c^2}\right)\label{effF1} \, , \label{effF2}
\end{eqnarray}
where the source can be viewed as a perfect fluid with effective energy density and pressure terms given by
\begin{eqnarray}
 \reff &=& \av{\r} - \frac{\Q}{16 \pi G}-\frac{c^2\av{^{(3)}\!R}}{16 \pi G}\\
  \peff &=& - \frac{c^2\Q}{16 \pi G}+\frac{c^4\av{^{(3)}\!R}}{48 \pi G}\,.
\end{eqnarray}
obeying the continuity equation
\begin{equation} \label{effcont}
\dot{\r}_{\D}^{eff} +3\frac{\da}{a_{\D}}\left(\r_{\D}^{eff} + \frac{P_{\D}^{eff}}{c^2}\right)=0\,,
\end{equation}
where 
\begin{equation} \label{curvav}
 \av{^{(3)}\!R} = \frac{1}{\mathcal{V}_{\mathcal{D}}}\int_{\mathcal{D}}{^{(3)}\!R\sqrt{h}~d^{3}q}
\end{equation}  
is the average spatial curvature and we have introduced the kinematical back-reaction
\begin{equation} \label{defQ}
 \mathcal{Q}_{\mathcal{\D}}=\frac{2}{3}\av{\left(\t - \av{\t}\right)^{2}}-2\av{\Sigma^{2}}\,.
\end{equation}

Consistency of Eq.~(\ref{effcont}) with mass conservation, Eq.~(\ref{masscons}), which can be re-written as 
\begin{equation} 
\dot{\r}_{\D} +3\frac{\da}{a_{\D}}\r_{\D}= 0\,,
\end{equation}
requires that the kinematical back-reaction and the mean spatial curvature 
satisfy the integrability condition
\begin{equation} \label{CNint}
\left(a_{\D}^{6}\Q\right)^{\centerdot}+c^2a_{\D}^{4}\left(a_{\D}^{2} \av{^{(3)}\!R}\right)^{\centerdot}=0\,.
\end{equation} 
We remark here that such a condition is a genuinely general relativistic effect, which has no analogue in Newtonian gravity,
since the curvature $\,^{(3)}\!R$ of comoving hypersurfaces vanishes identically in the Newtonian limit. Indeed, in the Newtonian 
limit the variance of the expansion rate and the shear combine to give a total derivative, so $\Q$ reduces to a pure boundary 
term \cite{buchertNEWTdust}, which becomes negligible if the averaging is performed over a volume of the order of the Hubble volume, 
as should be the case if one is interested in the back-reaction of inhomogeneities on the background expansion rate. 
It was shown in Ref.~\cite{KMR} that in order for the back-reaction of cosmological inhomogeneities 
to drive acceleration, the kinematical back-reaction should be $\mathcal{Q}_\mathcal{D} >  4\pi G \av{\varrho}$. 
 
There is another important feature of Eq.~(\ref{CNint}) which is worth pointing out, namely that a factor $c^2$ 
multiplies the mean spatial curvature term, implying that the Newtonian kinematical back-reaction 
$\overline{\mathcal{Q}}_{\mathcal{D}}$ couples with the PN mean spatial curvature $ \av{^{(3)}\!R}^{PN}$, the PN kinematical back-reaction 
${\mathcal{Q}}^{PN}_{\mathcal{D}}$ couples with the post-post-Newtonian mean spatial curvature $ \av{^{(3)}\!R}^{PPN}$ and so on.

By performing the conformal rescaling of the metric $h_{\a\b}$, the velocity-gradient tensor splits into a FRW and a peculiar term (recall~(\ref{rescdec}) )
\begin{equation}
 \t^\a_\b=\frac{1}{a}\left(\vartheta^\a_\b+\frac{2}{\tau}\d^\a_\b\right),
\end{equation}  
where $a$ is the scale factor of the underlying FRW background and we use here the conformal time coordinate. In terms of the peculiar quantities, the kinematical backreaction becomes 
\begin{equation} \label{Qpecul}
\mathcal{Q}_{\mathcal{\D}}=\frac{2}{3a^2}\left(\av{\vartheta^2} - \av{\vartheta}^{2}\right)-\frac{2}{a^2}\av{\ss^{2}} \,.
\end{equation}
Explicitly, for our one-dimensional Newtonian background we have $\overline{\ss}^2=\overline{\vartheta}^2/3$, thus Eq.~(\ref{Qpecul}) becomes
\begin{equation} \label{backnewt}
\overline{\mathcal{Q}}_{\mathcal{D}} =-\frac{2}{3a^2}\av{\overline{\vartheta}}^2\,, 
\end{equation}
where $a$ is our Einstein-de Sitter scale-factor, $\overline{\vartheta}= - \tau \du^2\varphi/ (3 (1-\tau^2\du^2\varphi/6)) $ and $\av{\dots}$ is our 
Newtonian average 
\begin{equation} \label{newtav}
\av{\dots} = \frac{a^3}{\overline{\mathcal{V}}_\mathcal{D}} \int_\mathcal{D} \dots \left(1-\tau^2\du^2\varphi/6\right) d^3q  \;,
\end{equation} 
where $\overline{\mathcal{V}}_\mathcal{D} = a^3 \int_\mathcal{D} \left(1-\tau^2\du^2\varphi/6\right) d^3q$. 
Recalling that 
\begin{equation}
 \frac{\overline{\vartheta}}{a}=\partial_{r_1}\overline{v}_1\,,
\end{equation} 
where $r_\a$ are proper Eulerian coordinates $r_\a=ax_\a$, $\overline{v}_1 \equiv - \tau \partial_1 \varphi/3$ is the Newtonian peculiar velocity in the direction $r_1$, we can re-write equation~(\ref{backnewt}) in the form
\begin{equation}
 \overline{\mathcal{Q}}_{\mathcal{D}} =-\frac{2}{3} \left(\av{\partial_{r_1}\overline{v}_1}^E \right)^2=0\, \;,
\end{equation} 
where we introduced the Eulerian average $\av{ \dots }^E \equiv (1/\int_{\mathcal D} d^3 r) \int_{\mathcal D} \dots d^3 r $.
Note that the last equality follows from the fact that $\av{\partial_{r_1}\overline{v}_1}^E = 0$, by construction (see Ref.~\cite{buchertNEWTdust}).
It follows that any analysis of back-reaction based on the Newtonian approximation is irrelevant. In addition, first and second order calculations
based on the fully relativistic perturbation theory are not sufficient for evaluating back-reaction. This means that fully
non-linear sub-Hubble perturbations must be taken into account. For these reasons, we use our non-linear PN 
metric to estimate the back-reaction term~(\ref{Qpecul}) in the PN approximation. 

In order to calculate the PN kinematical back-reaction $\Q^{PN}$, we consider the kinematical quantities related to the PN peculiar velocity-gradient tensor 
\begin{equation}
 \vartheta^\a_\b=\overline{\vartheta}^\a_\b+\frac{1}{2c^2}w^{\a'}_\b\,.
\end{equation} 
In full generality, without any assumption about the Newtonian background, the PN expansion of the average integrals in~(\ref{Qpecul}) leads to 
\begin{eqnarray} \label{backpn}
 \Q^{PN} &=&\frac{1}{3a^2}\av{\overline{\vartheta}^2 w}+\frac{2}{3a^2}\av{\overline{\vartheta}
 w'}-\frac{1}{3a^2}\av{\overline{\vartheta}^2}\av{w}-\frac{2}{3a^2}\av{\overline{\vartheta}}\av{w}'+   \\ \nonumber 
&&-\frac{1}{a^2}\av{\overline{\sigma}^2 w}-\frac{1}{a^2}\av{\left(\overline{\vartheta}^\a_\b 
w^{\b'}_\a-\frac{1}{3}\overline{\vartheta}w'\right)}+\frac{1}{a^2}\av{\overline{\sigma}^2}\av{w}\,,
\end{eqnarray}
where with $\left\langle \dots \right\rangle_\mathcal{D}$ 
we indicate the Newtonian average (PN corrections in the averaging procedure would yield higher-order terms).

Recalling that in our case $\overline{\ss}^2=\overline{\vartheta}^2/3$, equation~(\ref{backpn}) reduces to
\begin{eqnarray}
\label{qpn}
  \Q^{PN} &=&\frac{2}{a^2} \av{\overline{\vartheta}w^{2'}_2}-\frac{2}{3a^2}\av{\overline{\vartheta}}\av{w}' \nonumber\\
  & =&\frac{2}{a^2} \av{\overline{\vartheta}w^{2'}_{2}}\,,
\end{eqnarray}
where we used the commutation rule for $w$
\begin{equation} \label{commw}
\av{w}'-\av{w'}=\av{w\overline{\vartheta}}-\av{w}\av{\overline{\vartheta}} 
\end{equation}
and the last equality in Eq.~(\ref{qpn}) follows from the previous Newtonian calculation, having neglected the last term, which involves $\av{\overline{\vartheta}}$. 
Finally, using our Newtonian average~(\ref{newtav}) and our solution for $w^{2'}_{2}$, we find
\begin{equation}
 \Q^{PN} =-\frac{10 a\tau^2}{81\overline{\mathcal{V}}_\mathcal{D}} \int_\mathcal{D} \du (\du\varphi)^3 d^3q\,.
\end{equation} 
We can alternatively express this result in terms of Newtonian peculiar velocity $\overline{v}_1$. We find
\begin{equation}
 \Q^{PN} =\frac{10}{3 a\tau}\av{\partial_{r_1}\overline{v}_1^3}^E\,,
\end{equation} 
which indeed indicates a negligible PN correction.\\
Finally, using the PN spatial curvature
\begin{equation}
\label{PNSpatcurv}
^{(3)}\!R^{PN}= \frac{(20/3)\du^2\varphi}{a^2\left(1-\tau^2\du^2\varphi/6\right)}
\end{equation} 
the PN expansion of Eq.~(\ref{curvav}) leads to
\begin{equation}
 \av{^{(3)}\!R}^{PN}=\frac{20 a}{3\overline{\mathcal{V}}_\mathcal{D}}\int_\mathcal{D}  \du^2\varphi d^3q\,. 
\end{equation} 
It is then clear that, in the case of planar dynamics, both the kinematical back-reaction and the mean spatial curvature reduce 
to purely boundary terms \footnote{Note that these quantities can be seen as boundary terms both in Lagrangian and Eulerian coordinates, 
since -- at the Newtonian level -- $(d F(q_1) / dq^1) dq^1 = (d F(q_1(x_1,\tau) /dx) dx$, for any function $F(q_1)$.}    
and, as a consequence, they cannot lead to acceleration. Also interesting is the fact that $ \av{^{(3)}\!R}^{PN} \propto a^{-2}$, which provides a self-consistent 
solution of the integrability condition Eq.~(\ref{CNint}), for vanishing $\mathcal{Q}_\D$.

\subsection{Post-Newtonian corrections to the average scale-factor and expansion rate}

An alternative approach to back-reaction is that of computing directly the PN contribution to the average scale-factor and expansion rate.
Indeed, using the definition  $\ad=\left(\mathcal{V}_\D/\mathcal{V}_{\D_0}\right)^{1/3}$ , we can calculate the PN expansion of the average scale-factor, 
written as
\begin{equation}
 \ad=\overline{a}_{\mathcal{D}}+\frac{1}{c^2}\ad^{PN} \;. 
\end{equation}
For the Newtonian term, we have for the volume
\begin{equation}
 \overline{\mathcal{V}}_\D= \int_{\D} a^3 \left(1-\frac{\tau^2}{6}\du^2\varphi\right)d^3q\,,
\end{equation}  
thus, neglecting the boundary term, we obtain
\begin{equation}
 \overline{a}_{\mathcal{D}}=\frac{\tau^2}{\tau_0^2}\,,
\end{equation}
which is just the Einstein de Sitter scale factor $a(\tau)$. 

For the PN term we find
\begin{equation}
 \ad^{PN}=\frac{1}{6}a\left(\av{w} -\langle w_0\rangle_{\mathcal{D}_0}\right)\,,
\end{equation}
where $\left\langle \dots\right\rangle_\mathcal{D}$ is the Newtonian average and our solution gives 
\begin{equation}
w=\frac{-\frac{20}{63} \tau^4 \du^2\varphi\left(\du\varphi\right)^2 +\frac{5}{3}\tau^2\left[\left(4\left(a_{NL}-1\right) + 
1\right)\left(\du\varphi\right)^2+\left(4\left(a_{NL}-1\right)+2\right)\varphi\du^2\varphi\right] -60\varphi}{6-\tau^{2}\du^{2}\varphi}\,. 
\end{equation}
Explicitly, we have
\begin{eqnarray}
\ad^{PN}&=&\frac{a^4}{6\overline{\mathcal{V}}_\D}\int_{\D}\Biggl\{-\frac{10}{189}\du^2\varphi(\du\varphi)^2\tau^4 
+\frac{5}{18}\tau^2\left[\left(4\left(a_{NL}-1\right)+1\right)\left(\du\varphi\right)^2+ \right. \\ \nonumber
&& + \hphantom{\biggl\{}\left. \left(4\left(a_{NL}-1\right)+2\right)\varphi\du^2\varphi\right]  \Biggr\} \, d^3q \,,
\end{eqnarray}
which can be written as
\begin{equation}
\ad^{PN}=\frac{a^4}{6\overline{\mathcal{V}}_\D}\int_{\D}\left(-\frac{5}{18}\tau^2\left(\du\varphi\right)^2\right) \,d^3q
\end{equation} 
up to negligible boundary terms. The average scale-factor then becomes
\begin{equation}
 \ad = a\left( 1 - \frac{5}{12c^2} \av{\left(1+ \overline{\delta} \right) \overline{v}_1^2}  \right),
\end{equation}
which indicates a negligible PN correction.  \\
The PN contribution to the average expansion rate $\av{\Theta}=({3/a}){\ad' / \ad}$ is obtained by the expansion of its very definition, leading to
\begin{equation}
\av{\Theta}=3 H+\frac{1}{2ac^2}\av{w}'\,,
\end{equation}
where $H(t)=2/(a\tau) $ is the Hubble expansion rate of our Einstein-de Sitter background. 
Neglecting boundary terms, it is straightforward to obtain \footnote{If we expand Eq.~(\ref{avexprate}) up to second-order in perturbation theory, we find quantitative agreement with Eq. (41) in Ref.~\cite{KMNR}. We checked that the different numerical factor is simply due to the different definition of $\av{\delta\theta}$ used in the two calculations.}
\begin{equation} \label{avexprate}
\av{\Theta} = 3H \left( 1 - \frac{5}{12c^2} \av{\left(1+ \overline{\delta} \right) \overline{v}_1^2} \right).
\end{equation}
This result again implies that the PN correction is fully negligible for plane parallel perturbations.

\section{Comparison with the Szekeres solution}

In this paper we have considered the evolution of an irrotational and collisionless fluid in General Relativity in the synchronous and 
comoving gauge. Following the fluid-flow approach, \cite{ellis71}, it is possible to alternatively describe our system in terms of the 
fluid properties of irrotational dust, i.e. mass density, volume-expansion and shear tensor, and the electric and magnetic parts 
of the Weyl tensor. In addition, in the special case of plane-parallel dynamics considered here, the magnetic Weyl tensor vanishes identically, 
thus leading to the so-called {\em silent universe} case \cite{silent,silentprl,silentapj}, described by the set of equations
\begin{eqnarray}
\dot{\r}&=& -\t\r \\
\dot{\t}&=&-6\Sigma^{2}-\frac{1}{3}\t^{2}-4 \pi G \r   \\                                                                       
\dot{\Sigma}&=&\Sigma^2-\frac{2}{3}\t\Sigma-E \\
\dot{E}&=& -3E \Sigma- \t E -4 \pi G \r\Sigma \,,                                                                                                                                                                       
\end{eqnarray}
where $\t$ is the trace of velocity-gradient tensor, $\Sigma$ and $E$ are the eigenvalues of the shear tensor and of the electric Weyl tensor
\begin{equation}
 E^\a_\b = \frac{1}{3} \delta^\a_\b \left( 
\t^\mu_\nu \t^\nu_\mu - \t^2 \right) + \t
\t^\a_\b - \t^\a_\gamma \t^\gamma_\b 
+ c^2 \left( \,^{(3)}\!R^\a_\b - \frac{1}{3} \delta^\a_\b \,^{(3)}\!R \right)
\end{equation} 
in the directions $q_2$ and $q_3$. 
In this framework Croudace {\it et al.} \cite{croudace} obtain what they refere to as 
{\it relativistic Zel'dovich solution}, a sub-case of the exact solutions by Szekeres, \cite{szekeres}. This solution is made more appealing by the recent growing interest 
on the Szekeres metric in the general framework of studying inhomogeneous cosmologies as possible alternatives to FRW (see, e.g. Refs.~\cite{ishak1} and ~\cite{brunisz} and Refs. therein). 
Ref.~\cite{croudace} considers the relativistic 
evolution equations of silent universes. In the special case of local planar 
symmetry, i.e.  $\lambda_2=\lambda_3=0$, where $\lambda_i$ are the eigenvalues of the matrix $\partial_\a \partial_\b \varphi$, 
the velocity-gradient tensor reads \cite{silent} 
\begin{equation}
 \Theta^\a_\b= \frac{\dot{a}}{a} \pmatrix{  \label{Theta}
1 - \frac{a\lambda_1}{1-a\lambda_1} & & \cr
& 1 & \cr
 & & 1 },
\end{equation} 
which corresponds to the Newtonian Zel'dovich solution.
Croudace {\it et el.} then compute the associated metric from
\begin{equation}
 \Theta^\a_\b= \frac{1}{2}h^{\a\ss}\dot{h}_{\b\ss}
\end{equation} 
via the time-time component of Einstein's equations, i.e. the energy constraint, that closes the relativistic fluid-flow equations, completely 
fixing the spatial dependence of the metric. They show that the resulting metric coincides with the Szekeres form
\begin{eqnarray}
 h_{22}&=&h_{33}=1 \\
 h_{11}&=&\left(d(q_\a)-a(t)c(q_1)\right)^2
\end{eqnarray}
with
\begin{equation}
 d(q_\a)=d_{\rm{in}}(q_1)-\frac{5}{9}c(q_1)((q_2)^2+(q_3)^2)-a(t)c(q_1)\,. 
\end{equation} 

This solution leaves the FRW expansion unperturbed in the directions $q_2$ and $q_3$. In fact, following Ref.~\cite{silent}, Croudace {\em et al.} discard 
the sub-leading PN trace part of the initial conditions~(\ref{ci}), thereby considering perturbations in the direction $q_1$ only. 
On the contrary, we kept the one-dimensional initial seed $\varphi(q_1)$ in all directions, thereby allowing for perturbations in the component $h_{22}$ and 
$h_{33}$ of the metric \textit{ab initio}. As a consequence, our solution for $h_{22}$ and $h_{33}$ changes with time, showing a non-linear dependence 
on the gravitational potential $\varphi$. Hence our (approximate) solution {\em cannot} be recast into the Szekeres form.

\section{Conclusions}

In this paper we have obtained a solution of Einstein's field equations describing the non-linear cosmological dynamics of irrotational dust in Lagrangian coordinates for the 
specific case of plane-parallel perturbations, which is exact up to first post-Newtonian order.  Our solution (Eq.~(\ref{soluzione})), which represents 
the post-Newtonian extension of the well-known Zel'dovich solution for plane-parallel cosmological dynamics in Newtonian gravity, 
can be useful in many respects. Here we have seen just one possible application, 
in the frame of averaging and back-reaction of cosmological inhomogeneities. In a forthcoming paper we will analyze the properties of photon geodesics in our metric, to study, for instance, 
the luminosity distance-redshift relation in a strongly inhomogeneous Universe such as the one described by our metric. 

Our results also allow us to study the final stages of plane-parallel collapse, which obviously leads to a shell-crossing, pancake-like singularity.
Caustic formation is considered the main limitation of the Lagrangian description, both in the Newtonian approximation and in 
General Relativity.  As well-known, caustics arise because several fluid elements coming from different positions may converge to the 
same Eulerian position, thus forming infinite density regions. 
Such a pathological behavior occurs in our case when $f =-1$, i.e. when $\tau^{2}\du^{2}\varphi/6=1$: at this time the determinants of both the Newtonian~(\ref{metricor}) and PN 
metric~(\ref{soluzione}) go to zero, while the density contrast at both the Newtonian, Eq.~(\ref{caustden}), and PN level, (\ref{densPN}), becomes infinite, as does the PN spatial curvature in Eq.~(\ref{PNSpatcurv}). 
The appearance of shell-crossing singularities can be understood as indicating the breakdown of the dust approximation, rather than the occurrence of a true physical 
singularity of the gravitational collapse. The formation of caustics appears as an artifact of the extrapolation of the pressure-less fluid approximation beyond the point at which 
pressure has become important. 
An important result of our analysis is that the divergence of the PN spatial curvature at caustic formation is completely eliminated by the spatial smoothing 
procedure. Indeed $\mathcal{R}^{PN}$ diverges like $(1+\overline{\d})$, which is exactly compensated by the square root of the spatial metric determinant $\propto 1/(1+\overline{\d})$.  
This very fact confirms that the instability found in Ref.~\cite{KMR}, using a gradient expansion technique, and in Ref.~\cite{notari}, using a different approximation scheme, 
cannot be interpreted as a consequence of a shell-crossing singularity, but really arises from the back-reaction of sub-Hubble modes.
  
Our results here bring both good and bad news for the back-reaction of cosmic inhomogeneities to represent a potentially viable alternative to dark energy at a fundamental level. 
Let us start with the bad news: as we have shown the kinematical back-reaction scalar remains a negligible boundary term also at the post-Newtonian level, so that no relevant back-reaction effect
is implied by our PN solution. We should stress, however, that such a result is clearly a consequence of our very specific (but analytically solvable) model, 
which relies on one-dimensional dynamics. It is a very reasonable hypothesis that such a result will be modified by a (more complex!) full 3D calculation; this will be the subject of a subsequent paper. 
The good news are: the very fact that the Lagrangian approach allows to obtain a quantitative estimate of back-reaction and the fact that the appearance of shell-crossing singularities 
does not lead to divergences in the average dynamics of inhomogeneous dust Universes.

Also interesting, in our opinion, is the fact that the PN metric depends on the primordial non-Gaussianity strength parameter $a_{\rm NL}$ (which is related to the best known non-Gaussianity parameter
$f_{\rm NL}$ by $f_{\rm NL} = (5/3) a_{\rm NL}$, for large $a_{\rm NL}$).  
Although in our final results for $\mathcal{Q}_{\mathcal{D}}^{PN}$, $\av{^{(3)}\!R}^{PN}$ and $a_{\mathcal{D}}^{PN}$ this quantity disappeared 
(in that it only enters in boundary terms), one can safely predict that this parameter will affect back-reaction quantities for general three-dimensional 
perturbations, hence suggesting the intriguing possibility of a potential quantitative link between primordial non-Gaussianity and back-reaction.
   
\acknowledgments
We would like to acknowledge Marco Bruni, Enzo Branchini and Thomas Buchert for useful discussions and our referee for helpful comments.

\providecommand{\href}[2]{#2}\begingroup\raggedright\endgroup

\end{document}